\documentclass[aps,prl,showpacs,preprintnumbers,twocolumn]{revtex4-1}
\usepackage{geometry}                % See geometry.pdf to learn the layout options. There are lots.
\usepackage{graphicx}
\usepackage{amssymb}
\usepackage{dsfont}
\usepackage{epstopdf}
\usepackage{color}

\usepackage{mathtools}
\usepackage{hyperref}

\definecolor{red}{rgb}{1,0,0}
\definecolor{blue}{rgb}{0,0,1}
\definecolor{black}{rgb}{0,0,0}

\newcommand{\p}{\partial}
\newcommand{\eq}[1]{\begin{align}#1\end{align}}

\newcommand{\GG}{\mathcal{G}}

\newcommand{\start}{S}

\newcommand{\PP}{\mathbb{P}}

\newcommand{\TT}{\mathcal{T}}

\newcommand{\ed}{\epsilon_d}
\newcommand{\es}{\epsilon_s}

\usepackage{cancel}

\begin{document}
\title{Random Language Model}
\author{E. DeGiuli}
\affiliation{Institut de Physique Th\'eorique Philippe Meyer, \'Ecole Normale Sup\'erieure, \\ PSL University, Sorbonne Universit\'es, CNRS, 75005 Paris, France}
%\date{today}                     

\begin{abstract}
Many complex generative systems use languages to create structured objects. We consider a model of random languages, defined by weighted context-free grammars. As the distribution of grammar weights broadens, a transition is found from a random phase, in which sentences are indistinguishable from noise, to an organized phase in which nontrivial information is carried. This marks the emergence of deep structure in the language, and can be understood by a competition between energy and entropy. %Prospects for the 
\end{abstract}
 
 % possible referees: (T. Poschel, ..)
% J.P. Crutchfield, Professor of Physics, UC Davis, chaos (at) ucdavis (dot) edu 
% W. Bialek, John Archibald Wheeler/Battelle Professor in Physics, Princeton University, wbialek@princeton.edu
%M. Tegmark, Professor of Physics, MIT, tegmark@mit.edu

\maketitle
It is a remarkable fact that structures of the most astounding complexity can be encoded into sequences of digits from a finite alphabet. Indeed, the complexity of life is written in the genetic code, with alphabet $\{A,T,C,G\}$, proteins are coded from strings of 20 amino acids, and human-written text is composed in small, fixed alphabets. %, from Gibbon's `The Decline and Fall of the Roman Empire,' with 27 characters, to Feynman's Lectures on Physics, with, roughly, 80 characters. 
 This `infinite use of finite means' \cite{Von-Humboldt99} was formalized by Post and Chomsky with the notion of generative grammar \cite{Post43,Chomsky02}, and has been elaborated upon since, both by linguists and computer scientists \cite{Hopcroft07}. A generative grammar { consists of an alphabet of hidden symbols, an alphabet of observable symbols, and a set of rules, which allow certain combinations of symbols to be replaced by others. From an initial start symbol $S$, one progressively applies the rules until only observable symbols remain; any sentence produced this way is said to be `grammatical,' and the set of all such sentences is called the language of the grammar. %, operating by replacement, such that from 
%an initial start symbol $S$, one can produce a set of `grammatical' strings, called sentences. 
The sequence of rule applications is called a derivation. For example, the grammar $\{ S \to SS, S \to (S), S \to () \}$ has a single hidden symbol $S$ and two observable symbols, $($ and $)$, and produces the infinite set of all strings of well-formed parentheses. }%, which constitute the language of the grammar. 
 A simple derivation in this grammar is $S \to SS \to (S)S\to (())S\to (())()$. Besides their original use in linguistics, where the observable symbols are typically taken to be words, and grammars produce sentences (Fig 1a) \cite{Chomsky02,Chomsky14}, generative grammars have found application in manifold domains: in the secondary structure of RNA (Fig 1b) \cite{Searls02,Knudsen03}, in compiler design \cite{Hopcroft07}, in self-assembly \cite{Winfree99}, in protein sequence analysis \cite{Barton16}, and in quasicrystals \cite{Escudero97}, to name a few.

\begin{figure}[b!]
\includegraphics[width=\columnwidth]{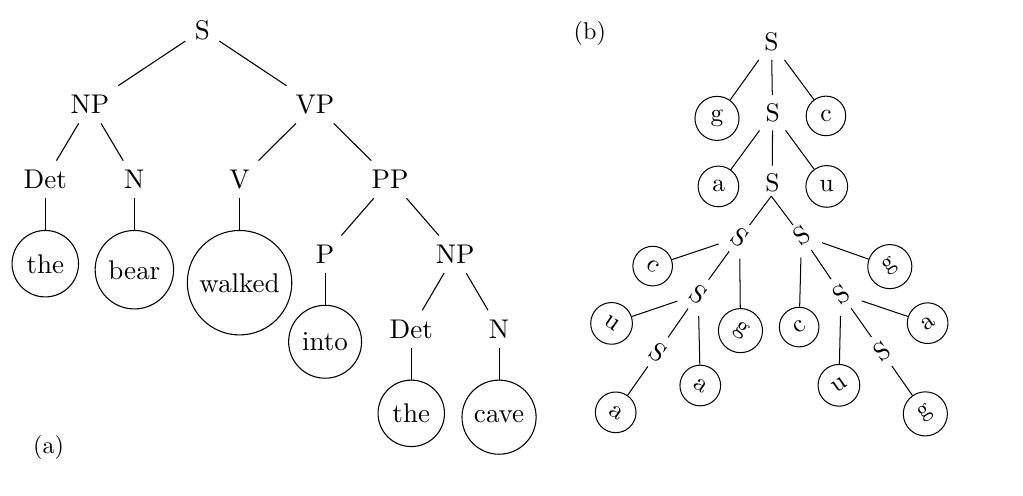}
\caption{ Illustrative derivation trees for (a) simple English sentence, and (b) RNA secondary structure (after \cite{Searls02}). The latter is a derivation of the sequence `gacuaagcugaguc' and shows its folded structure. Terminal symbols are encircled. 
}\label{fig1}
\end{figure} 

The complexity of a language is limited by conditions imposed on its grammar, as described by the Chomsky hierarchy, which, in increasing complexity, distinguishes regular, context-free, context-sensitive, and recursively enumerable grammars \cite{Nowak02}. Each class of grammar has a characteristic graphical structure of its derivations: regular grammars produce linear derivations, context-free grammars produce trees (Fig 1), and context-sensitive and recursively enumerable grammars produce more elaborate graphs. Associated with an increase in complexity is an increased difficulty of parsing \cite{Hopcroft07}. Because biological instantiations of grammars must have been discovered by evolution, there is a strong bias toward simpler grammars; %however, lacking internal structure in their derivations, regular grammars are quite limited, though useful in a stochastic setting, where they are known as hidden Markov models. 
we consider context-free grammars (CFGs), which are the lowest order of the Chomsky hierarchy that supports hierarchical structure. %, as well the simplest class that supports long-range information-theoretic correlations in strings \cite{Lin17}.
 %context-free grammars (CFGs) 
% play a special role. In what follows our discussion will be centered around them. %As recently shown by Lin and Tegmark, CFGs are also the simplest grammars that can lead to long-range information-theoretic correlations in strings. 

Despite their ubiquity in models of complex generative systems, grammars have hitherto played a minor role in physics, and most known results on grammars are theorems regarding worst-case behavior \footnote{{ For example, from Ref.\onlinecite{Hopcroft07}, Theorem 7.17 on the size of derivation trees, Theorem 7.31 on the conversion of an automaton to a CFG, and Theorem 7.32 on the complexity of conversion to Chomsky normal form (see below).}}, which need not represent the typical case. { Human languages show Zipf's law \cite{Zipf13,Cancho03,Corral15}, a power-law dependence of word frequency on its rank, and many sequences, including human text, show long-range information-theoretic correlations \cite{Ebeling94,Schurmann96,Lin17}, which can be created by a CFG \cite{Lin17}; but are these {\it typical} features of some ensemble of grammars? In this work we initiate this research program by proposing and simulating an ensemble of CFGs, so that grammars can be considered as physical systems \cite{Parisi99}. } %Each class in the Chomsky hierarchy defines an ensemble of grammars, which can be analyzed with the techniques of statistical physics. 
%: each class in the Chomsky hierarchy defines an ensemble of grammars, which can profitably be analyzed with the techniques of statistical physics. 
%We will show that, although initially appearing quite formidable, a Random Language Model can be reduced to a Potts model of unusual type. 
 We will find that CFGs possess two natural `temperature' scales that control grammar complexity, one at the surface interface, and another in the tree interior. As either of these temperatures is lowered, there is a phase transition, which corresponds to the emergence of nontrivial information propagation. We characterize this phase transition using results from simulations, and understand its location by a balance between energy and entropy.

{\bf Generative grammars:} A generative grammar is defined by an alphabet $\chi$ and a set of rules $\mathcal{R}$. The alphabet has $N$ hidden, `non-terminal' symbols $\chi_H$, and $T$ observable, `terminal' symbols $\chi_O$. The most general rule is of the form $a_1 a_2 \ldots a_n \to b_1 b_2 \ldots b_m$, where $a_i \in \chi_H, b_i \in \chi = \chi_H \cup \chi_O$. In a CFG the rules are specialized to the form $a_1 \to b_1 b_2 \ldots b_m$, and we will insist that $m\geq 1$, so that there is no `empty' string. Without loss of generality, we consider CFGs in Chomsky normal form, in which case all rules are of the form \cite{Hopcroft07} %\footnote{For simplicity we do not allow the empty string \cite{Hopcraft07}.}
$a \to b \;c$ or $a \to A$,
%\eq{
%a \to b \;c \qquad \mbox{or} \qquad a \to A,
%}
where $a,b,c \in \chi_H$ and $A \in \chi_O$.  Note that we may have $b=a$, or $b=c$, or $a=b=c$. Any derivation in Chomsky reduced form can be drawn on a binary tree. %We consider CFGs in this form. %; generalizations are discussed in the conclusion. 
 Beginning from the start symbol $\start \in \chi_H$, rules are applied until the string contains only observable symbols. Such a string is called a sentence. The set of all sentences is the language of the grammar. Given a string of observables $\mathcal{S} = A_1 \ldots A_\ell$ and a grammar $\GG$, one can ask whether there exists a derivation that produces $\mathcal{S}$ from the start symbol $\start$; if so, $\mathcal{S}$ is said to be grammatical. 

A formal grammar as defined above can only distinguish grammatical from ungrammatical sentences. A richer model is obtained by giving each rule a non-negative real valued weight. Such a weighted grammar is useful in applications, because weights can be continuously driven by a learning process, and can be used to define probabilities of parses. Moreover, a weighted grammar can be put into the Gibbs form, as shown below. %It it thus amenable to the tools of statistical mechanics. 
 For CFGs, to every rule of the form $a \to bc$ we assign a weight $M_{abc}$, and to every rule of the form $a \to A$ we assign a weight $O_{aA}$. %Furthermore, we give an extra weight $p_a$ when an internal rule $a \to bc$ is applied. 

Each candidate derivation of a sentence has two different types of degrees of freedom. There is the topology $\TT$ of the tree, namely the identity (terminal or non-terminal) of each node, as well as the variables, both terminal and non-terminal, on the nodes. We write $\Omega_\TT$ for the set of internal factors, i.e. factors of the form $a \to b c$, and $\p \Omega_\TT$ for the boundary factors, i.e. those associated to $a \to A$ rules. The number of boundary factors is written $\ell_\TT$, which is also the number of leaves. Since derivations are trees, the number of internal factors is $\ell_\TT-1$. We will write $\sigma$ for non-terminal symbols, and $o$ for terminals; these can be enumerated in an arbitrary way $1,\ldots,N$ and $1,\ldots,T$, respectively. Given $\TT$, we can write $\sigma_i$ for the value of the non-terminal on site $i$, and similarly $o_j$ for the terminal on site $j$. The number of $\sigma_i$ is $2\ell_\TT-1$, while the number of $o_j$ is $\ell_\TT$. We write $\GG$ for the pair $M,O$, $\sigma$ for $\{ \sigma_i \}$, and $o$ for $\{ o_t \}$.

To define a probability measure on derivations, it is convenient to factorize it into the part specifying $\TT$, and the remainder. In this way we separate the the tree shape from the influence of the grammar on variables. For a fixed $\TT$ the weight of a configuration is
\eq{ \label{w1}
W( \sigma,o | \TT, \GG) = \prod_{\alpha \in \Omega_\TT} M_{\sigma_{\alpha_1} \sigma_{\alpha_2} \sigma_{\alpha_3} } \prod_{\alpha \in \p\Omega_\TT} O_{\sigma_{\alpha_1} o_{\alpha_2} },
}
where each $\alpha=(\alpha_1,\alpha_2,\alpha_3)$ is a factor in the order $\sigma_{\alpha_1} \to \sigma_{\alpha_2} \sigma_{\alpha_3}$. Note that $M_{abc} \neq M_{acb}$ in general, thus the left and right branches are distinguished \footnote{Indeed if the left-right branches are not distinguished, CFGs do not have any more expressive power than regular grammars \cite{Esparza11}.}. We can write $W = e^{-E}$ with
\eq{ \label{E}
E = -\sum_{a,b,c} \pi_{abc}(\sigma) \log M_{abc} - \sum_{a,B} \rho_{aB}(\sigma,o) \log O_{aB}
}
where $\pi_{abc}$ is the number of times the rule $a\to bc$ appears in the configuration $\sigma$, and likewise $\rho_{aB}$ is the number of times the rule $a\to B$ appears. This defines a conditional probability measure on configurations $\PP(\sigma,o | \TT, \GG )~=~e^{-E(\sigma,o | \TT, \GG)}/ Z(\TT, \GG)$ 
%$\PP( \{\sigma_i, o_t \} | \TT, \GG )~=~e^{-E(\{\sigma_i, o_t \} | \TT, \GG)}/ Z(\TT, \GG)$ 
%\eq{
%\PP( \{\sigma_i, o_t \} | \TT, \GG ) = \frac{W( \{\sigma_i, o_t \} | \TT, \GG) }{Z(\TT, \GG)}
%}
where
\eq{
Z(\TT,\GG) = \sum_{\{ \sigma_i, o_t \} } e^{-E( \sigma, o | \TT, \GG)}.
}
All configurations have $S$ at the root node. For simplicity, in this work we consider as a model for the tree topology probability $\PP(\TT | \GG)  = W_{tree}/Z_{tree}$ with $W_{tree}(\TT) = p^{|\p \Omega_\TT|} (1-p)^{|\Omega_\TT|}$, 
%\eq{
%\frac{1}{Z_{tree}} \rho^{|\p \Omega_\TT|} (1-\rho)^{|\Omega_\TT|},
%}
where $p$ is the emission probability, the probability that a hidden node becomes an observable node. $p$  controls the size of trees; we will choose it such that the tree size distribution is cutoff above a length $\xi = 1000$. Some facts about the resulting binary trees are recorded in Supplementary Material \footnote{Supplementary Material includes details on binary trees, sampling methods, robustness in PCFG, differential entropies, and equation derivations, and Refs. \cite{Chib95,Flajolet09}.}. %, and the effects of a more general choice of $p_A$ are discussed in the conclusion.

%This assumes that $Z$ is finite. It is convenient to factorize this distribution into the part specifying the tree topology, and the remainder. We let $p_{a}$ be the relative probability that a site containing $a$ is the head node for an interior node, rather than a leaf node. The probability of a particular tree topology $\TT$ is then
%\eq{
%\PP(\TT | \GG) %& = \prod_{\alpha \in \Omega_\TT} \PP(\sigma_{\alpha_1} \to \mbox{non-terminal}) \prod_{\alpha \in \p\Omega_\TT} \PP(\sigma_{\alpha_1} \to \mbox{terminal}), \\
%& = \frac{1}{Z_{tree}} \prod_{\alpha \in \Omega_\TT} p_{\sigma_{\alpha_1}}
%}
%which in general depends on the hidden variables $\{\sigma_i\}$, despite the notation. 

A model with weights of the form \eqref{w1} is called a weighted CFG (WCFG). In the particular case where $1 = \sum_{b,c} M_{abc} = \sum_A O_{aA}$ for all $a$, it is easy to see that $M$ and $O$ are conditional probabilities: $M_{abc} = \PP(a \to bc \;|\; a \to \mbox{non-terminal})$ and $O_{aA} = \PP(a \to A \;|\; a \to \mbox{terminal})$. In this case the model is called a probabilistic CFG (PCFG). In the main text, we consider a weighted CFG, model W; in SI, we show that our results are robust in model P, a PCFG. %We will consider a model of each type in what follows: model W and model P. 
 There are tradeoffs between these models: model P is easier to sample, because it has $Z(\TT,\GG)=1$ from normalization of probability, and thus is factorized. But model W is more amenable to theory, since it is less constrained. 

%one would like a quantitative measure of grammaticality, so it is natural to consider a weighted CFG (WCFG) in which each rule is given a non-negative real valued weight. 

%: for example, in any real language, each sentence occurs with some frequency, 

%A special case of a weighted grammar is one in which the weights are normalized to be probabilities, that is, $ $. In this case sampling from the grammar is simple: one begins at the root of the tree, and progressively samples until a sentence is obtained. The price of this simplicity is that 
%stochastic grammar

{\bf Random Language Model: } Each grammar defines probabilities for sentences. To extract the universal properties of grammars, which do not depend on all details of $M$ and $O$, we need a measure on the space of grammars. %We now want to consider grammars themselves as random objects. 
 What is an appropriate measure? From Eq.\eqref{E}, $\log M$ and $\log O$ are analogous to coupling constants in statistical mechanics. A simple model is to assume a Gaussian distribution for these, so that $M$ and $O$ are lognormal. This can be motivated as follows: language evolution is a dynamical process, which must be slow in order for language to remain comprehensible at any given moment. If each $\log M_{abc}$ and $\log O_{aB}$ are the accumulation of independent, additive increments \cite{Sornette97}, these will lead to a lognormal.  
% Besides simplicity, this could be motivated if each $\log M_{abc}$ and $\log O_{aB}$ are the accumulation of independent, additive increments \cite{Sornette97}. These would correspond to dynamical increments in language evolution, which must be a slow process in order for language to remain comprehensible at any given moment. }
% A simple model is to consider grammar weights as the accumulation of many small multiplicative effects. }
 % We expect that grammar weights are the accumulation of many small multiplicative effects. If these are sufficiently independent, then they will lead to a lognormal distribution. 
 We define deep and surface sparsities as, respectively,
\eq{ \label{s1}
s_d = \frac{1}{N^3} \sum_{a,b,c} \log^2 \left[\frac{M_{abc}}{\overline{M}}\right], \;\; s_s =  \frac{1}{NT} \sum_{a,B} \log^2\left[\frac{O_{aB}}{\overline{O}} \right]
%s_d = \frac{1}{N^3} \sum_{a,b,c} \left|\log \frac{M_{abc}}{\overline{M}} \right|^2, \quad s_s =  \frac{1}{NT} \sum_{a,B} \left|\log \frac{O_{aB}}{\overline{O}} \right|^2
}
where $\overline{M}=1/N^2$ and $\overline{O}=1/T$ are the corresponding uniform probabilities; it is convenient to use this normalization even for model W where weights are not strictly normalized. A lognormal distribution of grammar weights is 
\eq{ \label{PPG}
\PP_G(M,O) & \equiv Z_G^{-1} \; J \; e^{-\epsilon_d s_d}  e^{- \epsilon_s s_s } %\notag \\
%& \qquad \; \prod_A \delta\big(1-\sum_{B,C} M_{ABC} \big) \delta\big(1-\sum_a O_{Aa} \big),
}
where $J = e^{-\sum_{a,b,c} \log M_{abc} - \sum_{a,B} \log O_{aB}}$, and the space of $M$ and $O$ is defined by appropriate normalization and positivity constraints. %$\delta_n$ imposes normalization of $M$ and $O$ probabilities. 
%In model W we impose only that $M$ and $O$ are positive; in model P we impose $\sum_{b,c} M_{abc}=1$ and $\sum_B O_{aB}=1$ for each $a$. 
 We define the Random Language Model as the ensemble of grammars drawn from Eq.\ref{PPG}. %; in analytical results this condition will be somewhat relaxed, as described below.

An alternative motivation of \eqref{PPG} is that this is the maximum-entropy measure when the grammar-averages % of $s_d$ and $s_s$ 
$\overline{s_d}$ and $\overline{s_s}$
 are constrained. $s_d$ and $s_s$ measure the density of rules about their respective median values $\overline{M}$ and $\overline{O}$. 
%Since $M$ and $O$ are probabilities, $s_d$ and $s_s$ are dominated by rules with small likelihood; they measure the density of rare rules. 
 When $s_d$ and $s_s$ are finite, all rules must have a finite probability: this reflects the fact that, given any finite amount of data, one can only put a lower bound on the probability of any particular rule. In model W the Lagrange multipliers $\epsilon_d$ and $\epsilon_s$ satisfy
\eq{ \label{s2}
\overline{s_d} = \frac{N^3}{2\epsilon_d}, \qquad \overline{s_s} = \frac{NT}{2\epsilon_s}.
%\overline{s_d} = \frac{N(N^2-1)}{2\epsilon_d}, \qquad \overline{s_s} = \frac{N(T-1)}{2\epsilon_s}.
%\overline{s_d} = |\log N^2|^2 + \frac{N(N^2-1)}{2\epsilon_d}, \qquad \overline{s_s} = |\log T|^2 + \frac{N(T-1)}{2\epsilon_s}.
}
When $\epsilon_d \to \infty$, $\overline{s_d} \to 0$, which is the value corresponding to a completely uniform deep grammar, that is, when for a non-terminal $a$, all rules $a \to b c$ have the same probability $1/N^2$. This is clearly the limit in which the grammar carries no information. As $\epsilon_d$ is lowered, $s_d$ increases, and the grammar carries more information. { In terms of how deterministic the rules are, $\epsilon_d$ plays the role of temperature, with random $\leftrightarrow$ hot and deterministic $\leftrightarrow$ cold; we will refer to it as the deep temperature. This analogy can also be seen formally: in SI, we show that if the energy $E$ is replaced by $\beta E$, then \eqref{s2} is replaced by $\overline{s_d} = \beta^2 N^3/(2\epsilon_d)$, such that lowering $\epsilon_d$ is equivalent to increasing $\beta$.} Similarly, $\epsilon_s$ controls information transmission at the surface; we call it the surface temperature.
%It can be shown that adding a drift term does not 
%We are interested in the extent to which languages carry information. It is instructive to consider a limit in which no information is carried. Suppose that, given a non-terminal $A$, all rules $A \to B,C$ have the same probability $1/N^2$. In this case, 

To investigate the role of $\ed$ on language structure, we sampled grammars from the RLM at fixed values $T=27,\es/(NT)=0.01$. Since the surface sparsity is large, there is already some simple structure at the surface; we will explore how deep structure emerges as $N$ and $\ed$ are varied. For each value of $N$ and $\ed$, we created 120 distinct grammars, from which we sample 200 sentences (see SI for more details). Altogether approximately 7200 distinct languages were constructed. 

\begin{figure}[t!]
\includegraphics[width=\columnwidth]{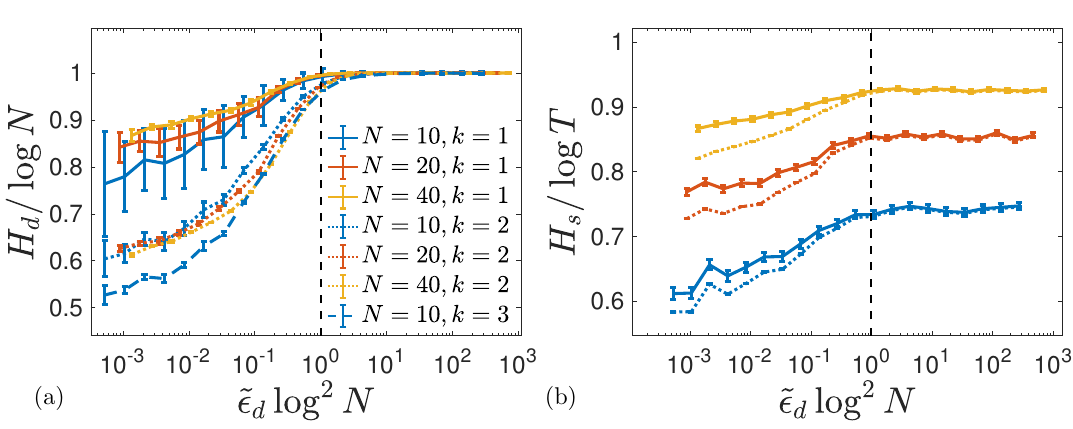}
\caption{ Shannon entropy of random CFGs as functions of $\tilde \epsilon_d = \epsilon_d / N^3$. (a) Block entropy of hidden configurations for indicated $k$ and $N$. (b) Block entropy of observed strings; symbols as in (a). The constant value for  $\epsilon_d > \epsilon_*$ depends on the surface temperature $\epsilon_s$. Bars indicate $20^{th}$ and $80^{th}$ percentiles.
}\label{fig2}
\end{figure} 

The information content of a grammar $\GG$ is naturally encoded by Shannon entropies. For a sequence $o_1,o_2,\ldots,o_k$ the Shannon block entropy rate is 
%rate of observed strings, 
\eq{ \label{S1}
H_s(\GG;k)=\frac{1}{k} \big\langle \log 1/\PP(o_1,o_2,\ldots,o_k | \GG) \big\rangle
%S_s(\GG) = \langle \log 1/\PP(o | \GG) \rangle.
}
For CFGs we can also consider the block entropy rate of deep configurations, 
\eq{ \label{S2}
H_d(\GG;k)=\frac{1}{k} \big\langle \log 1/\PP(\sigma_1,\sigma_2,\ldots,\sigma_k | \GG) \big\rangle
%S_d(\GG) = \langle \log 1/\PP(\sigma | \GG) \rangle,
}
where the symbols are taken from a (leftmost) derivation. 
\newcommand{\So}{\overline{H}}
 In both cases the ensemble average is taken with the actual probability of occurrence, $\PP(o|\GG)$ for $H_s$, and $\PP(\sigma|\GG)$ for $H_d$.  
% To accurately compute these entropies would require many sentences; here we consider instead the block entropies $S_s(\GG;k)=\frac{1}{k} \langle \log 1/\PP(o_1,o_2,\ldots,o_k | \GG) \rangle$ and $S_d(\GG;k)=\frac{1}{k}\langle \log 1/\PP(\sigma_1,\sigma_2,\ldots,\sigma_k | \GG) \rangle$, where the blocks $o_1,o_2,\ldots,o_k$ and $\sigma_1,\sigma_2,\ldots,\sigma_k$ are computed from adjacent symbols in derivations and sentences, respectively. In the limit $k\to \infty$, the block entropies converge to $S_s(\GG)$ and $S_d(\GG)$, respectively \cite{Schurmann96}.

\begin{figure}[t!]
\includegraphics[width=\columnwidth]{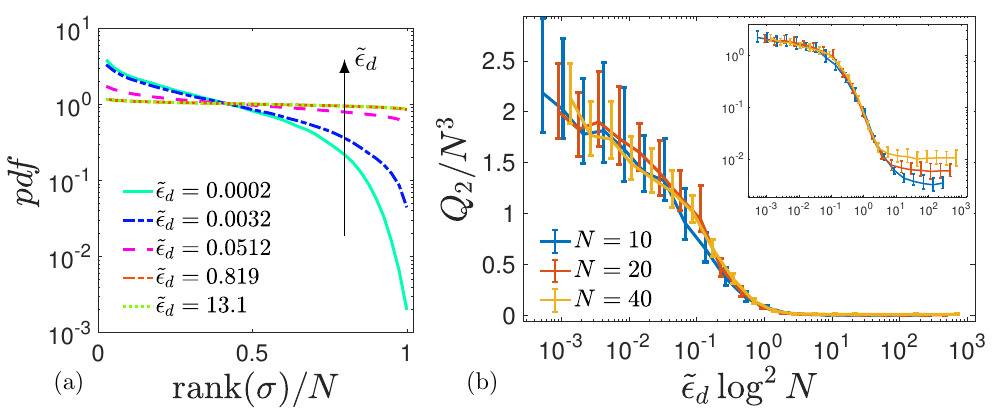}
\caption{ (a) Zipf plot of hidden symbols for $N=40$. Here $\tilde \epsilon_d = \epsilon_d/N^3$. (b) Order parameter $Q_2$, with bars indicating $20^{th}$ and $80^{th}$ percentile ranges over grammars at each parameter value. Inset: same plot in log-log axes.
}\label{fig3}
\end{figure}

The grammar averages $\So_d(k)$ and $\So_s(k)$ are shown in Fig. 2, for $k$ as indicated; here and in the following, the bars show the $20^{th}$ and $80^{th}$ percentiles, indicating the observable range of $H_d$ and $H_s$ over the ensemble of grammars \footnote{The error bars in measurements are then smaller by factor approximately $\sqrt{120} \sim 11$.}.  The dependence on $\ed$ is striking: for $\ed \gtrsim N^3/\log^2 N$, both $\So_s(1)$ and $\So_d(1)$ are flat. In this regime, $\So_d(1) \approx \log N$, indicating that although configurations strictly follow the rules of a WCFG, deep configurations are nearly indistinguishable from completely random configurations. However, at $\ed = \epsilon_* \approx N^3/\log^2 N$ there is a pronounced transition, and both entropies begin to drop. This transition corresponds to the emergence of deep structure. 

The first block entropy $H_d(\GG;1)$ measures information in the single-character distribution, while the differential entropies $\delta H_d(\GG;k) = (k+1) H_d(\GG;k+1) - k H_d(\GG;k)$ measure incremental information in the higher-order distributions \cite{Schurmann96}. The Shannon entropy rate including all correlations can either be obtained from $\lim_{k\to \infty} H_d(\GG;k)$, or from $\lim_{k\to \infty} \delta H_d(\GG;k)$. These coincide, but the latter converges faster \cite{Schurmann96}. In SI, we show that $\delta H_d(\GG;k)$, and thus the limiting rate, appears to collapse with $\tilde \epsilon_d \log N$. For all entropies the sample-to-sample fluctuations decrease rapidly with $k$, suggesting that the limiting rates are self-averaging. %$S_s(k\to \infty)$ and $S_d(k\to \infty)$ are self-averaging.
%are defined as the limits of $S_s(\GG;k)$ and $S_d(\GG;k)$ as $k \to \infty$ \cite{Schurmann96}. They can also be obtained from the limits of differential entropies $\delta S_s(\GG;k) = (k+1) S_s(\GG;k+1) - k S_s(\GG;k)$: $\lim_{k\to \infty} S_s(\GG;k) = \lim_{k\to \infty} \delta S_s(\GG;k)$ \cite{Schurmann96}. 

To further investigate the nature of the transition, we show in Fig. 3a a Zipf plot: the frequency of each symbol, arranged in decreasing order. Fig. 3a shows the Zipf plot for deep structure; the Zipf plot for surface structure is similar, but less dramatic (see SI). %and Fig 3b shows the Zipf plot for surface structure. 
 We see a sharp change at $\epsilon_*$: for $\ed > \epsilon_*$, the frequencies of hidden symbols are nearly uniform, while below $\epsilon_*$, the distribution is closer to exponential {( In SI, we show that a power-law regime for the observable symbols appears when $T$ is large)}. The permutation symmetry among hidden symbols is thus spontaneously broken at $\epsilon_*$.
 %The transition at $\epsilon_*$ is thus associated to symmetry breaking: the model has a full permutation symmetry among hidden symbols, which is spontaneously broken at $\epsilon_*$.  

What is the correct order parameter to describe this transition? The ferromagnetic order parameter is $m_r=\overline{\langle N \delta_{\sigma_i,r}-1 \rangle}$, where $i$ is a site. This does not show any signal of a transition, despite the fact that the start symbol explicitly breaks the replica symmetry. %\footnote{For a sentence of length $\ell$, if the start symbol only appears at the sentence head, then it will have a frequency $1/\ell$, which is greater than the uniform frequency $1/N$ for $\ell < N$. In our simulations we have {$\langle \ell \rangle \approx 10$}, and find that the start symbol appears with a frequency $\approx 0.4 N^{-1/2}$, smaller than $1/\ell$ for $N > 16$, so there is no indication of ferromagnetic order.}. 
 A more interesting choice is one of Edwards-Anderson type, such as $Q^{EA}_{rs} = \overline{ \langle N \delta_{\sigma_i,r} - 1 \rangle \langle N \delta_{\sigma_i,s} - 1 \rangle }$ where $r$ and $s$ label different sentences produced from the same grammar, and $\sigma_i$ is a specified site \cite{Gross85}. However, sentences produced by a CFG do not have fixed derivation trees, so we need to compare symbols in relative position. 
 %there is no sense in which sites can be identified in absolute position, so as to be compared between sentences. What can be defined is an order parameter with symbols in relative position. 
 For each interior rule $a\to bc$ we can define
\eq{ \label{Q1}
Q_{abc}(\GG) = \langle \delta_{\sigma_{\alpha_1},a} \big( N^2 \delta_{\sigma_{\alpha_2},b}\delta_{\sigma_{\alpha_3},c} -1 \big) \rangle,
}
averaged over all interior vertices $\alpha$, and averaged over derivations. Here $\sigma_{\alpha_1}$ is the head symbol at vertex $\alpha$, and $\sigma_{\alpha_2},\sigma_{\alpha_3}$ are the left and right symbols, respectively. $Q$ measures patterns in rule application at each branching of a derivation tree. %If $\sigma_{\alpha_1}$ is statistically independent of $\sigma_{\alpha_2}$ and $\sigma_{\alpha_3}$, then $Q_{ABC}$ will vanish. 
It is thus an order parameter for deep structure. Upon averaging over grammars in the absence of any fields, the permutation symmetry must be restored: 
%\eq{ \label{Q2}
$\overline{Q}_{abc} = q_0 + \delta_{ab} \; q_l + \delta_{ac} \; q_r + \delta_{bc} \; q_h + \delta_{ab}\delta_{ac} \; q_*$.
%}
As shown in SI, these components show a transition, but there is significant noise below $\epsilon_*$, despite there being $120$ replicas at each point.
%The component $q_0$ is plotted in Fig. 4a. This clearly shows the transition, but there is significant noise below $\epsilon_*$, despite there being $300$ replicas at each point. Similar results hold for $q_l,q_r,q_h$, and $q_*$. 
 Evidently, $Q_{abc}$ has large fluctuations below $\epsilon_*$. This suggests a definition
%\eq{ \label{Q21}
 $Q_2 \equiv \overline{ \sum_{a,b,c} Q_{abc}^2 },$
%}
plotted in Fig \ref{fig3}b. The signal is clear: on the large scale, $Q_2$ has a scaling form
%\eq{ \label{Q22}
%Q_2 \approx N^3 f\left( \frac{\epsilon_N \log^2 N}{N^3} \right)
 $Q_2 \approx N^3 f\left( \epsilon_N / \epsilon_* \right)$
%}
and is small above $\epsilon_*$. The scaling $Q_2 \sim N^3$ suggests that below the transition, all hidden symbols start to carry information in the deep structure. %The transition is 

{\bf Theory: } How can we gain some theoretical insight into the RLM? Consider the entropy of an observed string of length $\ell$, composed of $n$ sentences of length $\ell_k, \sum_k \ell_k = \ell$. The entropy of this string derives from 3 distinct combinatorial levels: %(i) the string can be separated into $p$ sentences in many different ways; 
(i) each sentence can be represented by a derivation tree with many different topologies; (ii) each derivation tree can host a variety of internal hidden variables; and (iii) given the hidden variables, the observed symbols can themselves vary. %By construction of distinct but related objects, each of these entropies can be evaluated.

Some scaling considerations are useful. 
%Consider a partition of the string into $p$ sentences of lengths $\ell_k$. In our model a tree of size $\ell_k$ has an intrinsic log-probability $(\ell_k-1) \log (1-p_E) + \ell_k \log p_E$, giving an effective entropic measure of the partition $S_{p} \sim (\ell-p) \log 1/(1-p_E) + \ell \log 1/p_E$. 
 Each derivation tree can have many topologies: the entropy of binary trees scales as $\ell_k \log 4$, so that the total tree entropy scales as $S_t \sim \ell \log 4$. Each derivation tree has $2 \ell_k - 1$ hidden variables, so that the total number of hidden DOF is $2 \ell - n$, and the corresponding deep entropy scales as $S_d \sim (2\ell-n) \log N$. Finally, the sentences have an entropy $S_o \sim \ell \log T$. 

% has an entropy $S_p = \sum_k \log \ell_k! - \log \ell \approx \sum_k \ell_k \log \frac{\ell_k}{\ell} \sim \ell \log \frac{\langle \ell \rangle}{\ell}$, where typical sentences are of length $\langle \ell \rangle$. 
%Each  Each tree topology carries an intrinsic log-probability $(\ell_k-1) \log (1-p_E) + \ell_k \log p_E$, giving an `energetic' contribution $E_{tp} \sim (\ell-p) \log (1-p_E) + \ell \log p_E$. 

We see that when typical sentences are of length $\langle \ell \rangle \gg 1$, so that $\ell-n \sim \ell$, these numbers are independent of partitioning, to leading order. %This is a special feature of trees, and would not hold in more general models. It corresponds to the fact that two trees can be combined into one by addition of a single site, connecting the roots. %provided the rule $\start \to \start \start$ has nonzero probability.
 For large $\langle \ell \rangle$ we get the scaling $S \sim \ell \log (4 N^2 T)$. 

This must be compared with the `energetic' terms $\log W_{tree} = (\ell-n) \log (1-p) + \ell \log p \sim -2 \ell \log 2$ for $p$ near $1/2$, and $E$, Eq.\eqref{E}. In $E$,
%\eq{
%\log W = \sum_{\alpha \in \Omega_\TT} \log M_{\sigma_{\alpha_1} \sigma_{\alpha_2} \sigma_{\alpha_3} }  + \sum_{\alpha \in \p\Omega_\TT} \log O_{\sigma_{\alpha_1} o_{\alpha_2} }
%}
% We can write
%\eq{
%\log W = \sum_{a,b,c} \pi_{abc}(\sigma) \log M_{abc} + \sum_{a,B} \rho_{aB}(\sigma,o) \log O_{aB}
%}
%where $\pi_{abc}$ is the number of times the rule $a\to bc$ appears in the configuration $\sigma$ (which may include multiple trees), and likewise $\rho_{aB}$ is the number of times the rule $a\to B$ appears.
%\eq{
%\pi_{abc}(\sigma) & = \sum_{\alpha \in \Omega_\TT} \delta_{\sigma_{\alpha_1},a} \delta_{\sigma_{\alpha_2},b} \delta_{\sigma_{\alpha_3},c} \\
%\rho_{aB}(\sigma,o) & = \sum_{\alpha \in \p \Omega_\TT} \delta_{\sigma_{\alpha_1},a} \delta_{o_{\alpha_2},B}
%}
%count the usage frequency of each rule in the grammar. 
 $\pi$ is positively correlated with $M$, since rules with a higher weight are more frequently used; hence we can obtain a simple scaling estimate $E \sim -N^3 \pi \log m -NT \rho \log o$ where $\pi$ is the mean value of $\pi_{abc}$, and $\log m$ is the value of a typical positive fluctuation of $\log M_{abc}$, and similarly for $O$. From the sum rules $\sum_{a,b,c} \pi_{abc} = |\Omega| = \ell - n$ and $\sum_{a,B} \rho_{aB} = |\p \Omega| = \ell$ we have $\pi = (\ell-n)/N^3, \rho = \ell/(NT)$. The mean value of $\log M_{abc}$ is $\log \overline{M}$% = -\log N^{2}$
, and the mean value of $\log O_{aB}$ is $\log \overline{O}$. % = - \log T$. 
 These contributions lead to a constant value of $E$. The positive fluctuations in $\log M$ and $\log O$ that couple to $E$ scale as $\sqrt{\frac{N^3}{2\epsilon_d}}$ and $\sqrt{\frac{NT}{2\epsilon_s}}$, respectively, leading to
\eq{
E \sim -\ell \sqrt{\frac{N^3}{2\epsilon_d}} - \ell \sqrt{\frac{NT}{2\epsilon_s}} + \mbox{const}
}
Combining this with $S$, the effective free energy $F = E-\log W_{tree} - S$ reflects a competition between energy and entropy. If we consider $N$ and $\epsilon_d$ as varying, then there is a scale $\epsilon_* = N^3/\log^2 N$ where the energetic fluctuations balance entropy. For $\epsilon_d \gg \epsilon_*$, the energy of a configuration is unimportant, and the grammar is thus irrelevant: the language produced by the WCFG must then be indistinguishable from random sequences, as found empirically above. In contrast, for $\epsilon_d \ll \epsilon_*$, the language reflects those sequences with high intrinsic weight, and their entropy is less important. The characteristic scale $\epsilon_*$ identified by these simple arguments agrees with that found empirically above, and locates the emergence of deep structure. 
%Thus at the coarsest level of description, the transition to the emergence of deep structure can be understood by a balance between entropy and energy.
 However, further work is needed to predict the behavior of $Q_2$, $H_s$, and $H_d$. %as well as the Shannon entropies $H_s$ and $H_d$.

{\bf Learning human languages: } Around 6000 languages are spoken around the world \cite{Baker08}; given fractured and highly sparse input, how does a child come to learn the precise syntax of one of these many languages? This question has a long history in linguistics and cognitive science \cite{Berwick11,Yang17}. One scenario for learning is known as the Principles and Parameters (P\&P) theory \cite{Chomsky93}. This posits that the child is biologically endowed with a general class of grammars, the `principles,' and by exposure to one particular language, fixes its syntax by setting some number of parameters, assumed to be binary. For example, the head-directionality parameter controls whether verbs come before or after objects, like English and Japanese, respectively. A vast effort has been devoted to mapping out the possible parameters of human languages
%, and their interrelations 
 \cite{Baker08,Shlonsky10}. The richness of the discovered %resulting 
  structure has been used as criticism of the approach \cite{Ramchand14}: if the child needs to set many parameters, then { do these all need to be innate? This would be a heavy evolutionary burden, and a challenge to efficient learning. } %the theory appears at odds with `poverty of the stimulus' arguments in favor of innate linguistic knowledge. 

The RLM can shed some light on this debate. First, since only 2 living human languages are known to possess syntax beyond CFG \footnote{Only Swiss-German and Bambara have confirmed features beyond CFG \cite{Culy85,Shieber85}.}, we consider WCFGs a valid starting point \footnote{Note also that some lexicalized models used for machine learning, such as \cite{Collins03}, are WCFGs with multi-indexed hidden variables.}. %, thus retaining the same basic structure.}. %thus giving $2/6000 = 0.000333$.} 
Following experimental work \cite{Yang17}, we picture the learning process as follows. Initially, the child does not know the rules of the grammar, so it begins with some small number of hidden symbols and assigns uniform values to the weights $M$ and $O$. To learn is to increase the likelihood of the grammar by adjusting the weights and adding new hidden symbols. %New hidden symbols are added when new data cannot be acceptably parsed. 
 As weights are driven away from uniform values, the temperatures $\epsilon_d$ and $\epsilon_s$ decrease. Eventually the transition to deep structure is encountered, and the grammar begins to carry information. 

In the absence of any bias, this transition would occur suddenly and dramatically, spontaneously breaking all $N^3$ directions in $M$ space simultaneously, as in Fig. 3b. However, in realistic child language learning, the child's environment acts as a field on this likelihood-ascent, and can cause the structure-emerging transitions to occur at different critical deep temperatures, depending on their coupling to the field. For example, %as the deep temperature is lowered, 
 a left-right symmetry breaking could correspond to setting the head directionality parameter. %Further in the structured phase, a weaker statistical symmetry could be broken, and so on. 

%As the deep temperature is lowered, the RLM is expected to spontaneously break any symmetries present: for example, a left-right symmetry breaking could correspond to setting the head directionality parameter. 
 
% We consider learning as a descent in the energy landscape, or as an ascent in likelihood landscape: initially, the child does not know the rules of the grammar, so it  
 
 Although this description is schematic, we insist that the various symmetry-breaking transitions, which could give rise to parameters, are emergent properties of the model. { Thus if there are indeed many parameters to be set, these do not all need to be innate: the child only needs the basic structure of a WCFG, and the rest is emergent. The P\&P theory is thus consistent with existence of many parameters. }%already implicit in the definition of the model, without any detailed additional information needed to be supplied. %An ascent in the energy landscape would encounter the various transitions. 
 If the RLM can be solved, by which we mean that the partition function $Z$ can be computed, then the series of symmetry-breaking transitions that occur in the presence of a field can be inferred, and a map of syntax in CFGs could be deduced. %the structure of the syntax of human languages could be deduced. 
 This is a tantalizing goal for future work.
 
 {\bf Conclusion: } We introduced a model of random languages, which captures the generative aspect of complex systems. 
 % could serve as a foundation for the physics of complex generative systems. 
 The model has a transition in parameter space that corresponds to the emergence of deep structure. Since the interaction is long-range, we expect that the RLM, or a variant, is exactly solvable. We hope that this will be clarified in the future.
  
\begin{acknowledgments}
This work benefited from discussions with C. Callan, J. Kurchan, G. Parisi, R. Monasson, G. Semerjian, P. Urbani, F. Zamponi, A. Zee, and Z. Zeravcic. %, and correspondence with N. Chomsky.
\end{acknowledgments}

\bibliographystyle{apsrev4-1}
\bibliography{../language}

%merlin.mbs apsrev4-1.bst 2010-07-25 4.21a (PWD, AO, DPC) hacked
%Control: key (0)
%Control: author (72) initials jnrlst
%Control: editor formatted (1) identically to author
%Control: production of article title (-1) disabled
%Control: page (0) single
%Control: year (1) truncated
%Control: production of eprint (0) enabled
\begin{thebibliography}{38}%
\makeatletter
\providecommand \@ifxundefined [1]{%
 \@ifx{#1\undefined}
}%
\providecommand \@ifnum [1]{%
 \ifnum #1\expandafter \@firstoftwo
 \else \expandafter \@secondoftwo
 \fi
}%
\providecommand \@ifx [1]{%
 \ifx #1\expandafter \@firstoftwo
 \else \expandafter \@secondoftwo
 \fi
}%
\providecommand \natexlab [1]{#1}%
\providecommand \enquote  [1]{``#1''}%
\providecommand \bibnamefont  [1]{#1}%
\providecommand \bibfnamefont [1]{#1}%
\providecommand \citenamefont [1]{#1}%
\providecommand \href@noop [0]{\@secondoftwo}%
\providecommand \href [0]{\begingroup \@sanitize@url \@href}%
\providecommand \@href[1]{\@@startlink{#1}\@@href}%
\providecommand \@@href[1]{\endgroup#1\@@endlink}%
\providecommand \@sanitize@url [0]{\catcode `\\12\catcode `\$12\catcode
  `\&12\catcode `\#12\catcode `\^12\catcode `\_12\catcode `\%12\relax}%
\providecommand \@@startlink[1]{}%
\providecommand \@@endlink[0]{}%
\providecommand \url  [0]{\begingroup\@sanitize@url \@url }%
\providecommand \@url [1]{\endgroup\@href {#1}{\urlprefix }}%
\providecommand \urlprefix  [0]{URL }%
\providecommand \Eprint [0]{\href }%
\providecommand \doibase [0]{http://dx.doi.org/}%
\providecommand \selectlanguage [0]{\@gobble}%
\providecommand \bibinfo  [0]{\@secondoftwo}%
\providecommand \bibfield  [0]{\@secondoftwo}%
\providecommand \translation [1]{[#1]}%
\providecommand \BibitemOpen [0]{}%
\providecommand \bibitemStop [0]{}%
\providecommand \bibitemNoStop [0]{.\EOS\space}%
\providecommand \EOS [0]{\spacefactor3000\relax}%
\providecommand \BibitemShut  [1]{\csname bibitem#1\endcsname}%
\let\auto@bib@innerbib\@empty
%</preamble>
\bibitem [{\citenamefont {Von~Humboldt}(1999)}]{Von-Humboldt99}%
  \BibitemOpen
  \bibfield  {author} {\bibinfo {author} {\bibfnamefont {W.}~\bibnamefont
  {Von~Humboldt}},\ }\href@noop {} {\emph {\bibinfo {title} {Humboldt:'On
  language': On the diversity of human language construction and its influence
  on the mental development of the human species}}}\ (\bibinfo  {publisher}
  {Cambridge University Press},\ \bibinfo {year} {1999})\BibitemShut {NoStop}%
\bibitem [{\citenamefont {Post}(1943)}]{Post43}%
  \BibitemOpen
  \bibfield  {author} {\bibinfo {author} {\bibfnamefont {E.~L.}\ \bibnamefont
  {Post}},\ }\href@noop {} {\bibfield  {journal} {\bibinfo  {journal} {American
  journal of mathematics}\ }\textbf {\bibinfo {volume} {65}},\ \bibinfo {pages}
  {197} (\bibinfo {year} {1943})}\BibitemShut {NoStop}%
\bibitem [{\citenamefont {Chomsky}(2002)}]{Chomsky02}%
  \BibitemOpen
  \bibfield  {author} {\bibinfo {author} {\bibfnamefont {N.}~\bibnamefont
  {Chomsky}},\ }\href@noop {} {\emph {\bibinfo {title} {Syntactic
  structures}}}\ (\bibinfo  {publisher} {Walter de Gruyter},\ \bibinfo
  {address} {Berlin},\ \bibinfo {year} {2002})\BibitemShut {NoStop}%
\bibitem [{\citenamefont {Hopcroft}\ \emph {et~al.}(2007)\citenamefont
  {Hopcroft}, \citenamefont {Motwani},\ and\ \citenamefont
  {Ullman}}]{Hopcroft07}%
  \BibitemOpen
  \bibfield  {author} {\bibinfo {author} {\bibfnamefont {J.~E.}\ \bibnamefont
  {Hopcroft}}, \bibinfo {author} {\bibfnamefont {R.}~\bibnamefont {Motwani}}, \
  and\ \bibinfo {author} {\bibfnamefont {J.~D.}\ \bibnamefont {Ullman}},\
  }\href@noop {} {\emph {\bibinfo {title} {Introduction to automata theory,
  languages, and computation}}},\ \bibinfo {edition} {3rd}\ ed.\ (\bibinfo
  {publisher} {Pearson},\ \bibinfo {address} {Boston, Ma},\ \bibinfo {year}
  {2007})\BibitemShut {NoStop}%
\bibitem [{\citenamefont {Chomsky}(2014)}]{Chomsky14}%
  \BibitemOpen
  \bibfield  {author} {\bibinfo {author} {\bibfnamefont {N.}~\bibnamefont
  {Chomsky}},\ }\href@noop {} {\emph {\bibinfo {title} {Aspects of the Theory
  of Syntax}}},\ Vol.~\bibinfo {volume} {11}\ (\bibinfo  {publisher} {MIT
  press},\ \bibinfo {address} {Cambridge},\ \bibinfo {year} {2014})\BibitemShut
  {NoStop}%
\bibitem [{\citenamefont {Searls}(2002)}]{Searls02}%
  \BibitemOpen
  \bibfield  {author} {\bibinfo {author} {\bibfnamefont {D.~B.}\ \bibnamefont
  {Searls}},\ }\href@noop {} {\bibfield  {journal} {\bibinfo  {journal}
  {Nature}\ }\textbf {\bibinfo {volume} {420}},\ \bibinfo {pages} {211}
  (\bibinfo {year} {2002})}\BibitemShut {NoStop}%
\bibitem [{\citenamefont {Knudsen}\ and\ \citenamefont
  {Hein}(2003)}]{Knudsen03}%
  \BibitemOpen
  \bibfield  {author} {\bibinfo {author} {\bibfnamefont {B.}~\bibnamefont
  {Knudsen}}\ and\ \bibinfo {author} {\bibfnamefont {J.}~\bibnamefont {Hein}},\
  }\href@noop {} {\bibfield  {journal} {\bibinfo  {journal} {Nucleic acids
  research}\ }\textbf {\bibinfo {volume} {31}},\ \bibinfo {pages} {3423}
  (\bibinfo {year} {2003})}\BibitemShut {NoStop}%
\bibitem [{\citenamefont {Winfree}\ \emph {et~al.}(1999)\citenamefont
  {Winfree}, \citenamefont {Yang},\ and\ \citenamefont {Seeman}}]{Winfree99}%
  \BibitemOpen
  \bibfield  {author} {\bibinfo {author} {\bibfnamefont {E.}~\bibnamefont
  {Winfree}}, \bibinfo {author} {\bibfnamefont {X.}~\bibnamefont {Yang}}, \
  and\ \bibinfo {author} {\bibfnamefont {N.~C.}\ \bibnamefont {Seeman}},\ }in\
  \href@noop {} {\emph {\bibinfo {booktitle} {DNA based computers II}}},\
  \bibinfo {series} {DIMACS series in discrete mathematics and theoretical
  computer science}, Vol.~\bibinfo {volume} {44}\ (\bibinfo  {publisher}
  {American Mathematical Soc.},\ \bibinfo {address} {Providence, R.I.},\
  \bibinfo {year} {1999})\ p.\ \bibinfo {pages} {191}\BibitemShut {NoStop}%
\bibitem [{\citenamefont {Barton}\ \emph {et~al.}(2016)\citenamefont {Barton},
  \citenamefont {Chakraborty}, \citenamefont {Cocco}, \citenamefont {Jacquin},\
  and\ \citenamefont {Monasson}}]{Barton16}%
  \BibitemOpen
  \bibfield  {author} {\bibinfo {author} {\bibfnamefont {J.~P.}\ \bibnamefont
  {Barton}}, \bibinfo {author} {\bibfnamefont {A.~K.}\ \bibnamefont
  {Chakraborty}}, \bibinfo {author} {\bibfnamefont {S.}~\bibnamefont {Cocco}},
  \bibinfo {author} {\bibfnamefont {H.}~\bibnamefont {Jacquin}}, \ and\
  \bibinfo {author} {\bibfnamefont {R.}~\bibnamefont {Monasson}},\ }\href@noop
  {} {\bibfield  {journal} {\bibinfo  {journal} {Journal of Statistical
  Physics}\ }\textbf {\bibinfo {volume} {162}},\ \bibinfo {pages} {1267}
  (\bibinfo {year} {2016})}\BibitemShut {NoStop}%
\bibitem [{\citenamefont {Escudero}(1997)}]{Escudero97}%
  \BibitemOpen
  \bibfield  {author} {\bibinfo {author} {\bibfnamefont {J.~G.}\ \bibnamefont
  {Escudero}},\ }in\ \href@noop {} {\emph {\bibinfo {booktitle} {Symmetries in
  Science IX}}}\ (\bibinfo  {publisher} {Springer},\ \bibinfo {address}
  {Boston},\ \bibinfo {year} {1997})\ pp.\ \bibinfo {pages}
  {139--152}\BibitemShut {NoStop}%
\bibitem [{\citenamefont {Nowak}\ \emph {et~al.}(2002)\citenamefont {Nowak},
  \citenamefont {Komarova},\ and\ \citenamefont {Niyogi}}]{Nowak02}%
  \BibitemOpen
  \bibfield  {author} {\bibinfo {author} {\bibfnamefont {M.~A.}\ \bibnamefont
  {Nowak}}, \bibinfo {author} {\bibfnamefont {N.~L.}\ \bibnamefont {Komarova}},
  \ and\ \bibinfo {author} {\bibfnamefont {P.}~\bibnamefont {Niyogi}},\
  }\href@noop {} {\bibfield  {journal} {\bibinfo  {journal} {Nature}\ }\textbf
  {\bibinfo {volume} {417}},\ \bibinfo {pages} {611} (\bibinfo {year}
  {2002})}\BibitemShut {NoStop}%
\bibitem [{Note1()}]{Note1}%
  \BibitemOpen
  \bibinfo {note} {{ For example, from Ref.\protect \rev@citealp {Hopcroft07},
  Theorem 7.17 on the size of derivation trees, Theorem 7.31 on the conversion
  of an automaton to a CFG, and Theorem 7.32 on the complexity of conversion to
  Chomsky normal form (see below).}}\BibitemShut {Stop}%
\bibitem [{\citenamefont {Zipf}(2013)}]{Zipf13}%
  \BibitemOpen
  \bibfield  {author} {\bibinfo {author} {\bibfnamefont {G.~K.}\ \bibnamefont
  {Zipf}},\ }\href@noop {} {\emph {\bibinfo {title} {The psycho-biology of
  language: An introduction to dynamic philology}}}\ (\bibinfo  {publisher}
  {Routledge},\ \bibinfo {address} {Milton Park},\ \bibinfo {year}
  {2013})\BibitemShut {NoStop}%
\bibitem [{\citenamefont {i~Cancho}\ and\ \citenamefont
  {Sol{\'e}}(2003)}]{Cancho03}%
  \BibitemOpen
  \bibfield  {author} {\bibinfo {author} {\bibfnamefont {R.~F.}\ \bibnamefont
  {i~Cancho}}\ and\ \bibinfo {author} {\bibfnamefont {R.~V.}\ \bibnamefont
  {Sol{\'e}}},\ }\href@noop {} {\bibfield  {journal} {\bibinfo  {journal}
  {Proceedings of the National Academy of Sciences}\ }\textbf {\bibinfo
  {volume} {100}},\ \bibinfo {pages} {788} (\bibinfo {year}
  {2003})}\BibitemShut {NoStop}%
\bibitem [{\citenamefont {Corral}\ \emph {et~al.}(2015)\citenamefont {Corral},
  \citenamefont {Boleda},\ and\ \citenamefont {Ferrer-i Cancho}}]{Corral15}%
  \BibitemOpen
  \bibfield  {author} {\bibinfo {author} {\bibfnamefont {A.}~\bibnamefont
  {Corral}}, \bibinfo {author} {\bibfnamefont {G.}~\bibnamefont {Boleda}}, \
  and\ \bibinfo {author} {\bibfnamefont {R.}~\bibnamefont {Ferrer-i Cancho}},\
  }\href@noop {} {\bibfield  {journal} {\bibinfo  {journal} {PloS one}\
  }\textbf {\bibinfo {volume} {10}},\ \bibinfo {pages} {e0129031} (\bibinfo
  {year} {2015})}\BibitemShut {NoStop}%
\bibitem [{\citenamefont {Ebeling}\ and\ \citenamefont
  {P\"oschel}(1994)}]{Ebeling94}%
  \BibitemOpen
  \bibfield  {author} {\bibinfo {author} {\bibfnamefont {W.}~\bibnamefont
  {Ebeling}}\ and\ \bibinfo {author} {\bibfnamefont {T.}~\bibnamefont
  {P\"oschel}},\ }\href@noop {} {\bibfield  {journal} {\bibinfo  {journal} {EPL
  (Europhysics Letters)}\ }\textbf {\bibinfo {volume} {26}},\ \bibinfo {pages}
  {241} (\bibinfo {year} {1994})}\BibitemShut {NoStop}%
\bibitem [{\citenamefont {Sch{\"u}rmann}\ and\ \citenamefont
  {Grassberger}(1996)}]{Schurmann96}%
  \BibitemOpen
  \bibfield  {author} {\bibinfo {author} {\bibfnamefont {T.}~\bibnamefont
  {Sch{\"u}rmann}}\ and\ \bibinfo {author} {\bibfnamefont {P.}~\bibnamefont
  {Grassberger}},\ }\href@noop {} {\bibfield  {journal} {\bibinfo  {journal}
  {Chaos: An Interdisciplinary Journal of Nonlinear Science}\ }\textbf
  {\bibinfo {volume} {6}},\ \bibinfo {pages} {414} (\bibinfo {year}
  {1996})}\BibitemShut {NoStop}%
\bibitem [{\citenamefont {Lin}\ and\ \citenamefont {Tegmark}(2017)}]{Lin17}%
  \BibitemOpen
  \bibfield  {author} {\bibinfo {author} {\bibfnamefont {H.~W.}\ \bibnamefont
  {Lin}}\ and\ \bibinfo {author} {\bibfnamefont {M.}~\bibnamefont {Tegmark}},\
  }\href@noop {} {\bibfield  {journal} {\bibinfo  {journal} {Entropy}\ }\textbf
  {\bibinfo {volume} {19}},\ \bibinfo {pages} {299} (\bibinfo {year}
  {2017})}\BibitemShut {NoStop}%
\bibitem [{\citenamefont {Parisi}(1999)}]{Parisi99}%
  \BibitemOpen
  \bibfield  {author} {\bibinfo {author} {\bibfnamefont {G.}~\bibnamefont
  {Parisi}},\ }\href@noop {} {\bibfield  {journal} {\bibinfo  {journal}
  {Physica A}\ }\textbf {\bibinfo {volume} {263}},\ \bibinfo {pages} {557}
  (\bibinfo {year} {1999})}\BibitemShut {NoStop}%
\bibitem [{Note2()}]{Note2}%
  \BibitemOpen
  \bibinfo {note} {Indeed if the left-right branches are not distinguished,
  CFGs do not have any more expressive power than regular grammars \cite
  {Esparza11}.}\BibitemShut {Stop}%
\bibitem [{Note3()}]{Note3}%
  \BibitemOpen
  \bibinfo {note} {Supplementary Material includes details on binary trees,
  sampling methods, robustness in PCFG, differential entropies, and equation
  derivations, and Refs. \cite {Chib95,Flajolet09}.}\BibitemShut {Stop}%
\bibitem [{\citenamefont {Sornette}\ and\ \citenamefont
  {Cont}(1997)}]{Sornette97}%
  \BibitemOpen
  \bibfield  {author} {\bibinfo {author} {\bibfnamefont {D.}~\bibnamefont
  {Sornette}}\ and\ \bibinfo {author} {\bibfnamefont {R.}~\bibnamefont
  {Cont}},\ }\href@noop {} {\bibfield  {journal} {\bibinfo  {journal} {Journal
  de Physique I}\ }\textbf {\bibinfo {volume} {7}},\ \bibinfo {pages} {431}
  (\bibinfo {year} {1997})}\BibitemShut {NoStop}%
\bibitem [{Note4()}]{Note4}%
  \BibitemOpen
  \bibinfo {note} {The error bars in measurements are then smaller by factor
  approximately $\protect \sqrt {120} \sim 11$.}\BibitemShut {Stop}%
\bibitem [{\citenamefont {Gross}\ \emph {et~al.}(1985)\citenamefont {Gross},
  \citenamefont {Kanter},\ and\ \citenamefont {Sompolinsky}}]{Gross85}%
  \BibitemOpen
  \bibfield  {author} {\bibinfo {author} {\bibfnamefont {D.}~\bibnamefont
  {Gross}}, \bibinfo {author} {\bibfnamefont {I.}~\bibnamefont {Kanter}}, \
  and\ \bibinfo {author} {\bibfnamefont {H.}~\bibnamefont {Sompolinsky}},\
  }\href@noop {} {\bibfield  {journal} {\bibinfo  {journal} {Physical review
  letters}\ }\textbf {\bibinfo {volume} {55}},\ \bibinfo {pages} {304}
  (\bibinfo {year} {1985})}\BibitemShut {NoStop}%
\bibitem [{\citenamefont {Baker}(2008)}]{Baker08}%
  \BibitemOpen
  \bibfield  {author} {\bibinfo {author} {\bibfnamefont {M.~C.}\ \bibnamefont
  {Baker}},\ }\href@noop {} {\emph {\bibinfo {title} {The atoms of language:
  The mind's hidden rules of grammar}}}\ (\bibinfo  {publisher} {Basic books},\
  \bibinfo {address} {New York},\ \bibinfo {year} {2008})\BibitemShut {NoStop}%
\bibitem [{\citenamefont {Berwick}\ \emph {et~al.}(2011)\citenamefont
  {Berwick}, \citenamefont {Pietroski}, \citenamefont {Yankama},\ and\
  \citenamefont {Chomsky}}]{Berwick11}%
  \BibitemOpen
  \bibfield  {author} {\bibinfo {author} {\bibfnamefont {R.~C.}\ \bibnamefont
  {Berwick}}, \bibinfo {author} {\bibfnamefont {P.}~\bibnamefont {Pietroski}},
  \bibinfo {author} {\bibfnamefont {B.}~\bibnamefont {Yankama}}, \ and\
  \bibinfo {author} {\bibfnamefont {N.}~\bibnamefont {Chomsky}},\ }\href@noop
  {} {\bibfield  {journal} {\bibinfo  {journal} {Cognitive Science}\ }\textbf
  {\bibinfo {volume} {35}},\ \bibinfo {pages} {1207} (\bibinfo {year}
  {2011})}\BibitemShut {NoStop}%
\bibitem [{\citenamefont {Yang}\ \emph {et~al.}(2017)\citenamefont {Yang},
  \citenamefont {Crain}, \citenamefont {Berwick}, \citenamefont {Chomsky},\
  and\ \citenamefont {Bolhuis}}]{Yang17}%
  \BibitemOpen
  \bibfield  {author} {\bibinfo {author} {\bibfnamefont {C.}~\bibnamefont
  {Yang}}, \bibinfo {author} {\bibfnamefont {S.}~\bibnamefont {Crain}},
  \bibinfo {author} {\bibfnamefont {R.~C.}\ \bibnamefont {Berwick}}, \bibinfo
  {author} {\bibfnamefont {N.}~\bibnamefont {Chomsky}}, \ and\ \bibinfo
  {author} {\bibfnamefont {J.~J.}\ \bibnamefont {Bolhuis}},\ }\href@noop {}
  {\bibfield  {journal} {\bibinfo  {journal} {Neuroscience and Biobehavioral
  Reviews}\ } (\bibinfo {year} {2017})}\BibitemShut {NoStop}%
\bibitem [{\citenamefont {Chomsky}(1993)}]{Chomsky93}%
  \BibitemOpen
  \bibfield  {author} {\bibinfo {author} {\bibfnamefont {N.}~\bibnamefont
  {Chomsky}},\ }\href@noop {} {\emph {\bibinfo {title} {Lectures on government
  and binding: The Pisa lectures}}},\ \bibinfo {number} {9}\ (\bibinfo
  {publisher} {Walter de Gruyter},\ \bibinfo {year} {1993})\BibitemShut
  {NoStop}%
\bibitem [{\citenamefont {Shlonsky}(2010)}]{Shlonsky10}%
  \BibitemOpen
  \bibfield  {author} {\bibinfo {author} {\bibfnamefont {U.}~\bibnamefont
  {Shlonsky}},\ }\href@noop {} {\bibfield  {journal} {\bibinfo  {journal}
  {Language and linguistics compass}\ }\textbf {\bibinfo {volume} {4}},\
  \bibinfo {pages} {417} (\bibinfo {year} {2010})}\BibitemShut {NoStop}%
\bibitem [{\citenamefont {Ramchand}\ and\ \citenamefont
  {Svenonius}(2014)}]{Ramchand14}%
  \BibitemOpen
  \bibfield  {author} {\bibinfo {author} {\bibfnamefont {G.}~\bibnamefont
  {Ramchand}}\ and\ \bibinfo {author} {\bibfnamefont {P.}~\bibnamefont
  {Svenonius}},\ }\href@noop {} {\bibfield  {journal} {\bibinfo  {journal}
  {Language Sciences}\ }\textbf {\bibinfo {volume} {46}},\ \bibinfo {pages}
  {152} (\bibinfo {year} {2014})}\BibitemShut {NoStop}%
\bibitem [{Note5()}]{Note5}%
  \BibitemOpen
  \bibinfo {note} {Only Swiss-German and Bambara have confirmed features beyond
  CFG \cite {Culy85,Shieber85}.}\BibitemShut {Stop}%
\bibitem [{Note6()}]{Note6}%
  \BibitemOpen
  \bibinfo {note} {Note also that some lexicalized models used for machine
  learning, such as \cite {Collins03}, are WCFGs with multi-indexed hidden
  variables.}\BibitemShut {Stop}%
\bibitem [{\citenamefont {Esparza}\ \emph {et~al.}(2011)\citenamefont
  {Esparza}, \citenamefont {Ganty}, \citenamefont {Kiefer},\ and\ \citenamefont
  {Luttenberger}}]{Esparza11}%
  \BibitemOpen
  \bibfield  {author} {\bibinfo {author} {\bibfnamefont {J.}~\bibnamefont
  {Esparza}}, \bibinfo {author} {\bibfnamefont {P.}~\bibnamefont {Ganty}},
  \bibinfo {author} {\bibfnamefont {S.}~\bibnamefont {Kiefer}}, \ and\ \bibinfo
  {author} {\bibfnamefont {M.}~\bibnamefont {Luttenberger}},\ }\href@noop {}
  {\bibfield  {journal} {\bibinfo  {journal} {Information Processing Letters}\
  }\textbf {\bibinfo {volume} {111}},\ \bibinfo {pages} {614} (\bibinfo {year}
  {2011})}\BibitemShut {NoStop}%
\bibitem [{\citenamefont {Chib}\ and\ \citenamefont
  {Greenberg}(1995)}]{Chib95}%
  \BibitemOpen
  \bibfield  {author} {\bibinfo {author} {\bibfnamefont {S.}~\bibnamefont
  {Chib}}\ and\ \bibinfo {author} {\bibfnamefont {E.}~\bibnamefont
  {Greenberg}},\ }\href@noop {} {\bibfield  {journal} {\bibinfo  {journal} {The
  american statistician}\ }\textbf {\bibinfo {volume} {49}},\ \bibinfo {pages}
  {327} (\bibinfo {year} {1995})}\BibitemShut {NoStop}%
\bibitem [{\citenamefont {Flajolet}\ and\ \citenamefont
  {Sedgewick}(2009)}]{Flajolet09}%
  \BibitemOpen
  \bibfield  {author} {\bibinfo {author} {\bibfnamefont {P.}~\bibnamefont
  {Flajolet}}\ and\ \bibinfo {author} {\bibfnamefont {R.}~\bibnamefont
  {Sedgewick}},\ }\href@noop {} {\emph {\bibinfo {title} {Analytic
  combinatorics}}}\ (\bibinfo  {publisher} {cambridge University press},\
  \bibinfo {year} {2009})\BibitemShut {NoStop}%
\bibitem [{\citenamefont {Culy}(1985)}]{Culy85}%
  \BibitemOpen
  \bibfield  {author} {\bibinfo {author} {\bibfnamefont {C.}~\bibnamefont
  {Culy}},\ }\href@noop {} {\bibfield  {journal} {\bibinfo  {journal}
  {Linguistics and Philosophy}\ }\textbf {\bibinfo {volume} {8}},\ \bibinfo
  {pages} {345} (\bibinfo {year} {1985})}\BibitemShut {NoStop}%
\bibitem [{\citenamefont {Shieber}(1985)}]{Shieber85}%
  \BibitemOpen
  \bibfield  {author} {\bibinfo {author} {\bibfnamefont {S.~M.}\ \bibnamefont
  {Shieber}},\ }in\ \href@noop {} {\emph {\bibinfo {booktitle} {Philosophy,
  Language, and Artificial Intelligence}}}\ (\bibinfo  {publisher} {Springer},\
  \bibinfo {year} {1985})\ pp.\ \bibinfo {pages} {79--89}\BibitemShut {NoStop}%
\bibitem [{\citenamefont {Collins}(2003)}]{Collins03}%
  \BibitemOpen
  \bibfield  {author} {\bibinfo {author} {\bibfnamefont {M.}~\bibnamefont
  {Collins}},\ }\href@noop {} {\bibfield  {journal} {\bibinfo  {journal}
  {Computational linguistics}\ }\textbf {\bibinfo {volume} {29}},\ \bibinfo
  {pages} {589} (\bibinfo {year} {2003})}\BibitemShut {NoStop}%
\end{thebibliography}%


%merlin.mbs apsrev4-1.bst 2010-07-25 4.21a (PWD, AO, DPC) hacked
%Control: key (0)
%Control: author (8) initials jnrlst
%Control: editor formatted (1) identically to author
%Control: production of article title (-1) disabled
%Control: page (0) single
%Control: year (1) truncated
%Control: production of eprint (0) enabled
\begin{thebibliography}{6}%
\makeatletter
\providecommand \@ifxundefined [1]{%
 \@ifx{#1\undefined}
}%
\providecommand \@ifnum [1]{%
 \ifnum #1\expandafter \@firstoftwo
 \else \expandafter \@secondoftwo
 \fi
}%
\providecommand \@ifx [1]{%
 \ifx #1\expandafter \@firstoftwo
 \else \expandafter \@secondoftwo
 \fi
}%
\providecommand \natexlab [1]{#1}%
\providecommand \enquote  [1]{``#1''}%
\providecommand \bibnamefont  [1]{#1}%
\providecommand \bibfnamefont [1]{#1}%
\providecommand \citenamefont [1]{#1}%
\providecommand \href@noop [0]{\@secondoftwo}%
\providecommand \href [0]{\begingroup \@sanitize@url \@href}%
\providecommand \@href[1]{\@@startlink{#1}\@@href}%
\providecommand \@@href[1]{\endgroup#1\@@endlink}%
\providecommand \@sanitize@url [0]{\catcode `\\12\catcode `\$12\catcode
  `\&12\catcode `\#12\catcode `\^12\catcode `\_12\catcode `\%12\relax}%
\providecommand \@@startlink[1]{}%
\providecommand \@@endlink[0]{}%
\providecommand \url  [0]{\begingroup\@sanitize@url \@url }%
\providecommand \@url [1]{\endgroup\@href {#1}{\urlprefix }}%
\providecommand \urlprefix  [0]{URL }%
\providecommand \Eprint [0]{\href }%
\providecommand \doibase [0]{http://dx.doi.org/}%
\providecommand \selectlanguage [0]{\@gobble}%
\providecommand \bibinfo  [0]{\@secondoftwo}%
\providecommand \bibfield  [0]{\@secondoftwo}%
\providecommand \translation [1]{[#1]}%
\providecommand \BibitemOpen [0]{}%
\providecommand \bibitemStop [0]{}%
\providecommand \bibitemNoStop [0]{.\EOS\space}%
\providecommand \EOS [0]{\spacefactor3000\relax}%
\providecommand \BibitemShut  [1]{\csname bibitem#1\endcsname}%
\let\auto@bib@innerbib\@empty
%</preamble>
\bibitem [{\citenamefont {Chib}\ and\ \citenamefont
  {Greenberg}(1995)}]{Chib95}%
  \BibitemOpen
  \bibfield  {author} {\bibinfo {author} {\bibfnamefont {S.}~\bibnamefont
  {Chib}}\ and\ \bibinfo {author} {\bibfnamefont {E.}~\bibnamefont
  {Greenberg}},\ }\href@noop {} {\bibfield  {journal} {\bibinfo  {journal} {The
  american statistician}\ }\textbf {\bibinfo {volume} {49}},\ \bibinfo {pages}
  {327} (\bibinfo {year} {1995})}\BibitemShut {NoStop}%
\bibitem [{\citenamefont {Sch{\"u}rmann}\ and\ \citenamefont
  {Grassberger}(1996)}]{Schurmann96}%
  \BibitemOpen
  \bibfield  {author} {\bibinfo {author} {\bibfnamefont {T.}~\bibnamefont
  {Sch{\"u}rmann}}\ and\ \bibinfo {author} {\bibfnamefont {P.}~\bibnamefont
  {Grassberger}},\ }\href@noop {} {\bibfield  {journal} {\bibinfo  {journal}
  {Chaos: An Interdisciplinary Journal of Nonlinear Science}\ }\textbf
  {\bibinfo {volume} {6}},\ \bibinfo {pages} {414} (\bibinfo {year}
  {1996})}\BibitemShut {NoStop}%
\bibitem [{\citenamefont {Flajolet}\ and\ \citenamefont
  {Sedgewick}(2009)}]{Flajolet09}%
  \BibitemOpen
  \bibfield  {author} {\bibinfo {author} {\bibfnamefont {P.}~\bibnamefont
  {Flajolet}}\ and\ \bibinfo {author} {\bibfnamefont {R.}~\bibnamefont
  {Sedgewick}},\ }\href@noop {} {\emph {\bibinfo {title} {Analytic
  combinatorics}}}\ (\bibinfo  {publisher} {cambridge University press},\
  \bibinfo {year} {2009})\BibitemShut {NoStop}%
\bibitem [{\citenamefont {Zipf}(2013)}]{Zipf13}%
  \BibitemOpen
  \bibfield  {author} {\bibinfo {author} {\bibfnamefont {G.~K.}\ \bibnamefont
  {Zipf}},\ }\href@noop {} {\emph {\bibinfo {title} {The psycho-biology of
  language: An introduction to dynamic philology}}}\ (\bibinfo  {publisher}
  {Routledge},\ \bibinfo {address} {Milton Park},\ \bibinfo {year}
  {2013})\BibitemShut {NoStop}%
\bibitem [{\citenamefont {i~Cancho}\ and\ \citenamefont
  {Sol{\'e}}(2003)}]{Cancho03}%
  \BibitemOpen
  \bibfield  {author} {\bibinfo {author} {\bibfnamefont {R.~F.}\ \bibnamefont
  {i~Cancho}}\ and\ \bibinfo {author} {\bibfnamefont {R.~V.}\ \bibnamefont
  {Sol{\'e}}},\ }\href@noop {} {\bibfield  {journal} {\bibinfo  {journal}
  {Proceedings of the National Academy of Sciences}\ }\textbf {\bibinfo
  {volume} {100}},\ \bibinfo {pages} {788} (\bibinfo {year}
  {2003})}\BibitemShut {NoStop}%
\bibitem [{\citenamefont {Corral}\ \emph {et~al.}(2015)\citenamefont {Corral},
  \citenamefont {Boleda},\ and\ \citenamefont {Ferrer-i Cancho}}]{Corral15}%
  \BibitemOpen
  \bibfield  {author} {\bibinfo {author} {\bibfnamefont {A.}~\bibnamefont
  {Corral}}, \bibinfo {author} {\bibfnamefont {G.}~\bibnamefont {Boleda}}, \
  and\ \bibinfo {author} {\bibfnamefont {R.}~\bibnamefont {Ferrer-i Cancho}},\
  }\href@noop {} {\bibfield  {journal} {\bibinfo  {journal} {PloS one}\
  }\textbf {\bibinfo {volume} {10}},\ \bibinfo {pages} {e0129031} (\bibinfo
  {year} {2015})}\BibitemShut {NoStop}%
\end{thebibliography}%
\end{document}

\title{Random Language Model -- Supplementary Information}
\author{E. DeGiuli}
\affiliation{Institut de Physique Th\'eorique Philippe Meyer, \'Ecole Normale Sup\'erieure, \\ PSL University, Sorbonne Universit\'es, CNRS, 75005 Paris, France}
%\date{today}    

\maketitle

{\bf Sampling Methods: } %In our model we first sample a tree topology $\TT$ according to the binomial model where, given a non-terminal, this node branches into two non-terminals with probability $1-p$, and becomes a terminal node with probability $p$. Given $\TT$, we then sample the node variables as follows. 
 We consider both WCFGs and PCFGs. Since their distribution is factorized, PCFGs are trivial to sample: we begin at the top of the tree with $\start$, and choose whether this branches into two non-terminals with probability $1-p$, or becomes a terminal node with probability $p$. $S$ is replaced by non-terminals or terminals according to the probabilities specified in $M$ and $O$, respectively. This process is then repeated, where we act by replacement on the left-most non-terminal (for CFGs the order of replacement does not affect the derivation), replacing non-terminals according to the probabilities $M$ or $O$; we continue until no non-terminals remain. It is possible that sequences grow unboundedly; we stop any that go beyond length 4000. This occurs for less than a fraction $10^{-4}$ of samples. It is clear that the emission probability $p$ could be absorbed into $M$ and $O$: the advantage of our model is that we strictly control the typical sentence length, so that this does not appear as a confounding variable in the analysis, and we avoid creation of infinite trees. 

WCFGs are sampled with the Metropolis-Hastings algorithm \cite{Chib95}. For a given WCFG, we define a related PCFG by $\tilde M_{abc} = M_{abc} / \sum_{b',c'} M_{ab'c'}$, $\tilde O_{aB} = O_{aB} / \sum_{B'} O_{aB'}$, for all $a$. This PCFG is used as a candidate-generating density. In the Metropolis-Hastings rule, the individual factors $M_{abc}, O_{aB}$ all cancel, leaving only the normalization factors from $\sum_{b',c'} M_{ab'c'}$, $\sum_{B'} O_{aB'}$, which have smaller fluctuations than the individual $M_{abc}, O_{aB}$. This ensures efficient sampling.

For model W, for each value of $N$ and $\ed$, we created 120 distinct grammars, from which we sample 200 sentences. We fix the emission probability $p$ so that the mean sentence length is $\langle \ell \rangle \approx 15$, and the total length of text sampled for each grammar is $\approx 3000$. 

Block entropies are computed using the method of Grassberger (Eq. 8 in \cite{Schurmann96}). Since a deep block entropy of order $k$ has a phase space with $N^k$ configurations, we can only effectively compute entropies for which $N^k \lesssim 3000$, and similarly surface block entropies can be computed for $T^k \lesssim 3000$. Shown entropies in the main text satisfy this bound.  They are absent from finite $\ell$ effects: we saw no dependence when entropies were computed using only the first half of the sampled sentences.

\begin{figure}[t!]
\includegraphics[width=\columnwidth]{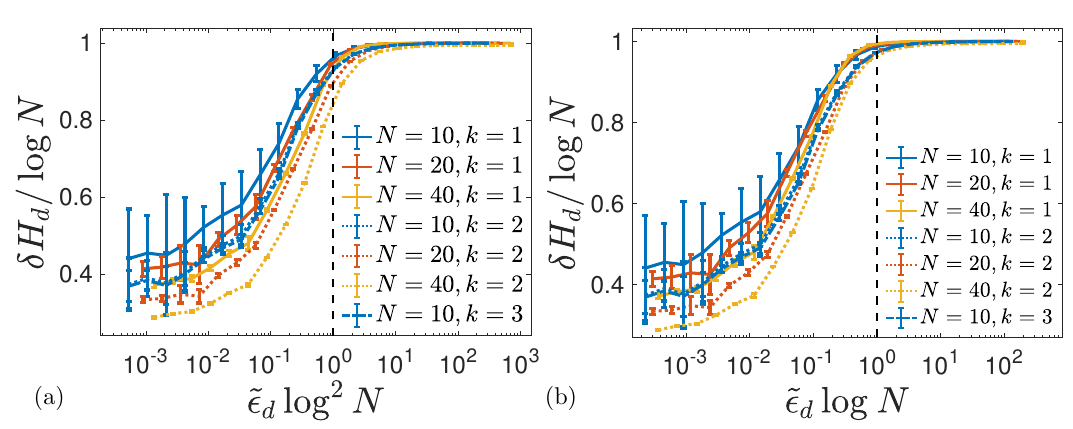}
\caption{ Differential entropy of hidden symbols of random WCFGs as functions of $\tilde \epsilon_d = \epsilon_d / N^3$, for indicated $k$ and $N$. (a) versus $\tilde \epsilon_d \log^2 N$; (b) versus $\tilde \epsilon_d \log N$. Bars indicate $20^{th}$ and $80^{th}$ percentiles.
}\label{figA1}
\end{figure} 
\begin{figure}[t!]
\includegraphics[width=\columnwidth]{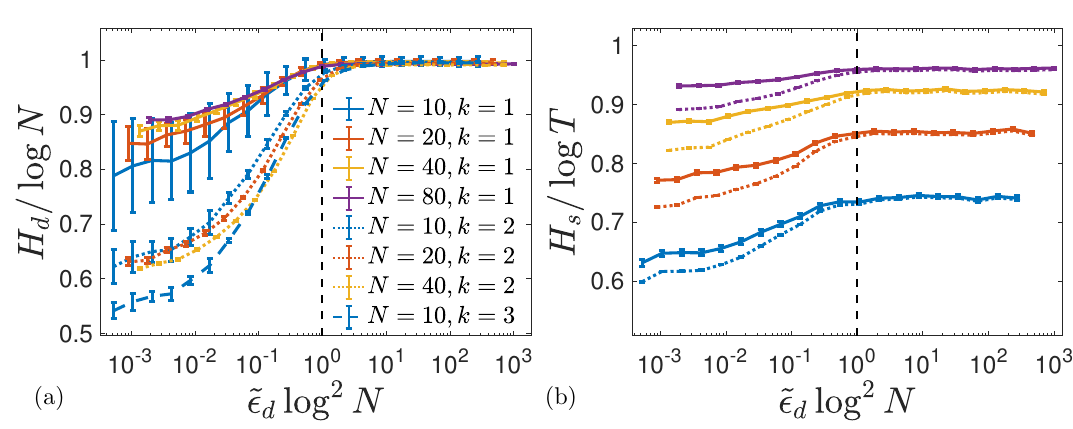}
\caption{ Shannon entropy of random PCFGs as functions of $\tilde \epsilon_d = \epsilon_d / N^3$. (a) Block entropy of hidden configurations for indicated $k$ and $N$. (b) Block entropy of observed strings. Symbols are as in (a), although a different subset of parameters is shown. The constant value for  $\epsilon_d > \epsilon_*$ depends on the surface temperature $\epsilon_s$. Bars indicate $20^{th}$ and $80^{th}$ percentiles.
}\label{figA2}
\end{figure}

{\bf Binary trees: } Our model for the probability of a tree topology $\TT$ is
%\eq{
%\PP(\TT | \GG) = \frac{1}{Z_{tree}} (\rho(1-\rho))^{|\Omega_\TT|},
%}
%where $|\Omega_\TT|$ is the number of internal factors in the tree. For a binary tree with $\ell$ leaves, $|\Omega_\TT| = \ell-1$. Since the number of boundary factors $|\p \Omega_\TT|$ is $\ell$, we can write this as
\eq{
\PP(\TT | \GG) = \frac{1}{Z_{tree}} (1-p)^{|\Omega_\TT|} p^{|\p \Omega_\TT|},
}
For a binary tree with $\ell$ leaves, $|\Omega_\TT| = \ell-1$ and $|\p \Omega_\TT| = \ell$. 
%making it clear that $\rho$ is the probability that a non-terminal becomes a terminal, that is, the emission probability. 
The number of binary trees with $\ell$ leaves is given\cite{Flajolet09} by the Catalan number 
\eq{
C_{\ell-1} = \frac{1}{\ell} \binom{2(\ell-1)}{\ell-1} \sim \frac{4^\ell}{4 \sqrt{\pi} (\ell-1)^{3/2}}
}
leading to
\eq{
Z_{tree} = p \sum_{\ell=1}^\infty C_{\ell-1} (p(1-p))^{\ell-1} = \frac{2p}{1+|2p-1|}, %= \frac{2p}{1 + \sqrt{1-4 p(1-p)}},
} 
where we used the result for the generating function of the Catalan numbers. As expected there is a singularity at $p=1/2$. Let us write $p = 1/2 + \epsilon$ with $\epsilon>0$. Then $Z_{tree}=1$ and one can show that the distribution of tree sizes follows
\eq{
\PP(\ell) \propto \frac{e^{-\ell/\xi}}{\ell^{3/2}},
}
where $\xi = 1/(4\epsilon^2)$. In our numerical results we have set $\xi=1000$. 

\begin{figure*}[bh!]
\includegraphics[width=\textwidth]{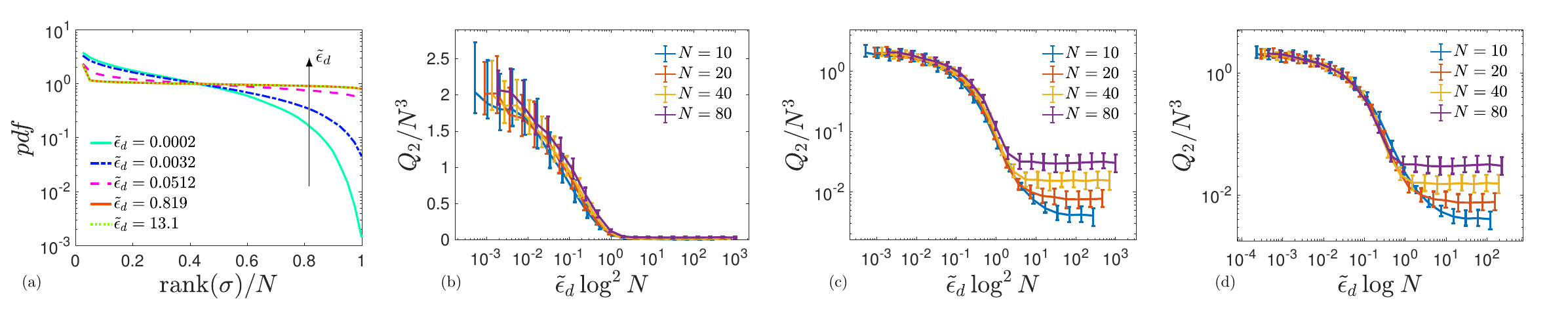}
\caption{ Results for model P. (a) Zipf plot of hidden symbols for $N=40$. Here $\tilde \epsilon_d = \epsilon_d/N^3$. (b-d) Order parameter $Q_2$, with bars indicating $20^{th}$ and $80^{th}$ percentile ranges over grammars at each parameter value. (d) shows a plot vs $\log N$ rather than $\log^2 N$. The collapse is better at large $N$ but worse at small $N$.
}\label{figA3}
\end{figure*} 

{
{\bf Scaling symmetry with temperature: } We can generalize the model by adding a bias $\beta$ to the energy, such that the probability of a configuration is $\PP(\sigma,o | \TT, \GG )~=~e^{-\beta E(\sigma,o | \TT, \GG)}/ Z(\TT, \GG)$. This is equivalent to replacing $M_{abc}$ by $M_{abc}^\beta$ and $O_{aB}$ by $O_{aB}^\beta$. If we also rescale $\overline{M}$ as $\overline{M}' = \overline{M}^\beta$, and  $\overline{O}' = \overline{O}^\beta$, then the sparsities rescale as $s_d' = \beta^2 s_d, \; s_s' = \beta^2 s_s$. Thus the rescaled equations of state are
\eq{ \label{as2}
\overline{s_d}' = \frac{\beta^2 N^3}{2\epsilon_d}, \qquad \overline{s_s}' = \frac{\beta^2 NT}{2\epsilon_s}.
}
Hence decreasing $\epsilon_d$ and $\epsilon_s$ is equivalent to increasing $\beta$, as well as rescaling the median values $\overline{M}$ and $\overline{O}$. It can be shown that grammar-averages of moments of the partition function $\overline{Z(\TT,\GG)^m}$ also have this scaling property. 
}
{\bf Differential entropy: } Deep differential entropies are defined from block entropies by 
\eq{
\delta H_d(\GG;k) = (k+1) H_d(\GG;k+1) - k H_d(\GG;k)
}
To examine the behavior as $k$ is increased, we performed additional simulations over an ensemble of 30 WCFGs at each $\epsilon_d$ and $N$, with 5000 sampled sentences per grammar. The resulting $\delta H_d$ are plotted in Fig. 1, versus (a) $\tilde \epsilon_d \log^2 N$, and (b) $\tilde \epsilon_d \log N$, where $\tilde \epsilon_d = \epsilon_d / N^3$. Collapse is better in the latter case, suggesting that the limiting rate obtained in the limit $k \to \infty$ depends on this reduced variable. 

%{ The approach to the $k\to \infty$ asymptote indicates whether derivations have long-range correlations, or not \cite{Ebeling94,Bialek01}. Long-range correlations were found in \cite{Ebeling94}, with }

{\bf Model P: } We consider PCFGs with $T=27, \epsilon_s/NT = 0.01$, and $N=10,20,40,80$ with varying $\epsilon_d$. For each parameter set, 300 grammars were constructed, and 200 sentences were sampled. Altogether 24000 languages were considered. Results are shown in Figs. \ref{figA2},\ref{figA3}. All qualitative behavior is identical to that found for model W. Scaling collapses are the same, but the choice of $n$ in $\log^n N$ may differ: as shown in Fig.\ref{figA3}cd, for large $N$ the value $n=1$ collapses better than $n=2$, which is optimal for small $N$. 
 
%\vfill

{
{\bf Zipf plots: } Zipf plots for the observed symbols are shown in Fig.\ref{figA3}. At large $\epsilon_d$, these are not flat, unlike the Zipf plots for hidden symbols, because we have fixed $\epsilon_s/(NT)=0.01$. Still, as the transition at $\epsilon_d=\epsilon_*$ is encountered, the distributions broaden. In WCFGs with a larger observable alphabet, a power-law regime emerges, as shown in Fig.\ref{figA4b}. For rank $R$ greater than $N$, the exponent is near $-2$, while at smaller $R$, the behavior depends on $N$, $T$, and $\epsilon_d$. The small $R$ regime also depends on $\tilde\epsilon_s$ (not shown). For comparison, a slope $-1$, commonly observed in human language \cite{Zipf13,Cancho03,Corral15}, is also shown. It has been proposed that this exponent arises from constraints of efficient communication \cite{Zipf13,Cancho03}. }
\begin{figure*}[t!]
\includegraphics[width=1.3\columnwidth]{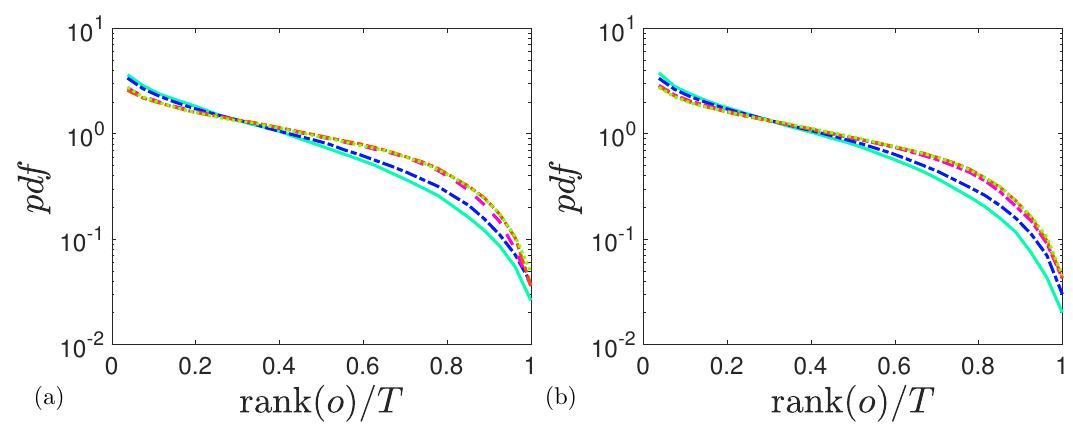}
\caption{ Zipf plots. Frequency of observed symbols, in decreasing order. (a) WCFG (b) PCFG. In both cases $N=40$, and labels are as in Fig. \ref{figA3}a. 
}\label{figA4}
\end{figure*} 
\begin{figure*}[h!]
\includegraphics[width=1.3\columnwidth]{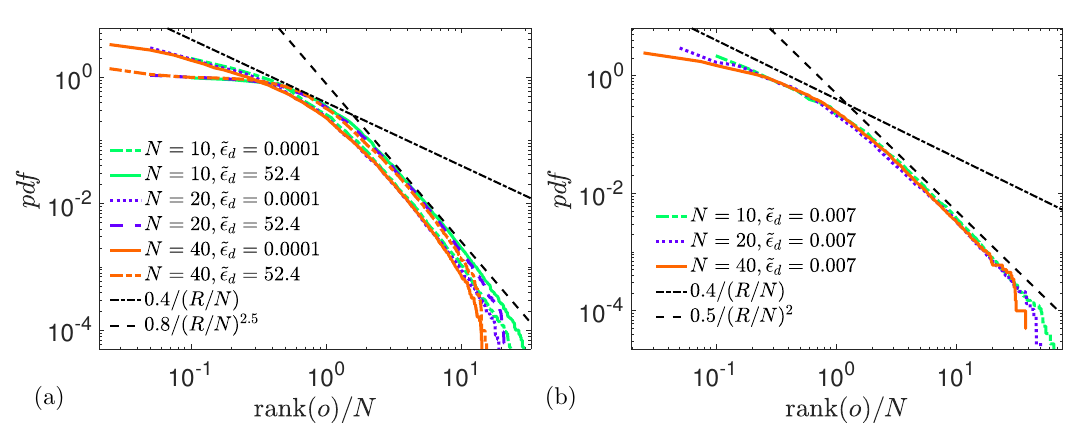}
\caption{ Zipf plot in WCFG with (a) $T=1000$ and (b) $T=10000$. Frequency of observed symbols, in decreasing order, together with indicated power laws as a function of the rank $R$. 
}\label{figA4b}
\end{figure*} 

{\bf Equation-of-state: } In model W, the grammar partition function is
\eq{
Z_G = \int dM \int dO \; J \; e^{-\epsilon_d s_d}  e^{- \epsilon_s s_s }
}
One easily sees that 
\eq{
\overline{s_d} = -\frac{\p \log Z_G}{\p \epsilon_d}, \;\;  \overline{s_s} = -\frac{\p \log Z_G}{\p \epsilon_s}
}
After a change of variable $m_{abc} = \log M_{abc}/\overline{M}, o_{aB} = \log O_{aB}/\overline{O}$, $Z_G$ is Gaussian and we find
\eq{
Z_G = \overline{J} \left( \frac{\pi N^3}{\epsilon_d} \right)^{\half N^3} \left( \frac{\pi NT}{\epsilon_s} \right)^{\half NT} 
}
leading to 
\eq{
\overline{s_d} = \frac{N^3}{2\epsilon_d}, \;\;\overline{s_s} = \frac{NT}{2\epsilon_s}
}
In model P, we have a normalization $\delta-$function that can be integrated using its Fourier transform. The final result cannot be put into a very explicit form. %However, it is straightforward to treat a model variation in which the sum of the log-weights is strictly imposed (model L), for each head symbol $a$. Since this constrains one DOF per head symbol, it leads to equations of state of the form
%\eq{
%\overline{s_d} = \frac{N (N^2-1)}{2\epsilon_d}, \;\;\overline{s_s} = \frac{N(T-1)}{2\epsilon_s}
%}
%We find that this model behaves very similarly to models W and P, as functions of $\epsilon_d$ and $\epsilon_s$.

\begin{figure*}[b!]
\includegraphics[width=1.3\columnwidth]{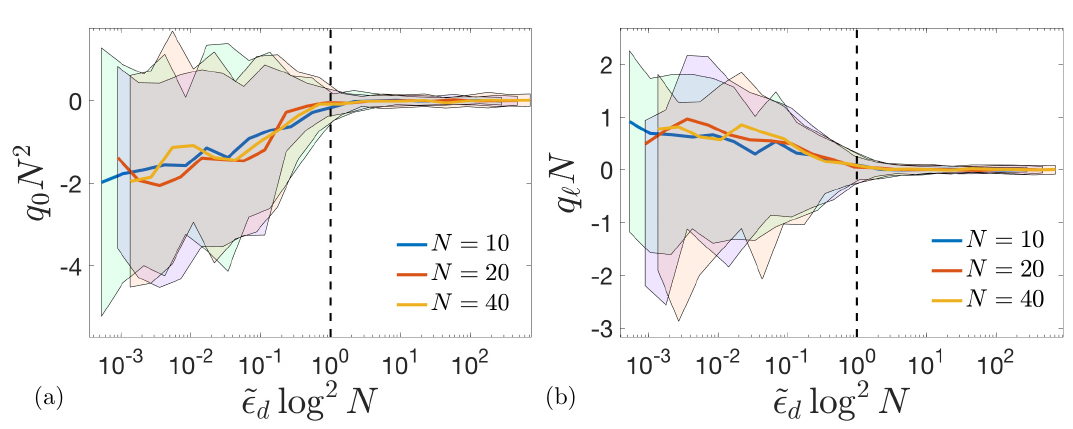}
\caption{ Grammar-averages of scalars (a) $q_0$ and (b) $q_l$ as functions of $\tilde \epsilon_d = \epsilon_d / N^3$. Bands indicate $20^{th}$ and $80^{th}$ percentiles.
}\label{figA5}
\end{figure*}

{\bf Deep structure symmetries: } In the absence of fields, the order parameter $Q_{abc}$ has a grammar average that must respect permutation symmetry. It then is of the form
\eq{ %\label{AppQ2}
\overline{Q}_{abc} = q_0 + \delta_{ab} \; q_l + \delta_{ac} \; q_r + \delta_{bc} \; q_h + \delta_{ab}\delta_{ac} \; q_*,
}
where $N^2 q_0 + N (q_l+q_r+q_h) + q_* =0$ from $\sum_{b,c} Q_{abc}=0$. $q_0$ and $q_l$ are plotted, for model W, in Fig.\ref{figA5}. ($q_r$ behaves the same as $q_l$, by symmetry. $q_*$ and $q_h$ display even larger fluctuations.) The bands show $20^{th}$ and $80^{th}$ percentiles over the 120 grammars sampled at each parameter value. We see that these quantities display wild fluctuations, much larger than the quantity $Q_2$. 

\vfill

\bibliography{../language}